\documentclass[aps,amsmath,amssymb,nofootinbib,showpacs,notitlepage,byrevtex,showkeys,prd,
]{revtex4-1}

\usepackage{graphicx}
\usepackage{subfigure}
\usepackage{bbm}
\usepackage{color}

\renewcommand{\vec}[1]{\boldsymbol{#1}}

\newcommand{\beq}{\begin{equation}}
\newcommand{\eeq}{\end{equation}}

\newcommand{\ii}{\mathrm{i}}
\newcommand{\dd}{\mathrm{d}}
\newcommand*{\da}[1][]{\mathop{\mathrm{d}\mkern-7mu\mathchar'26\mkern-1mu^{#1}}\mkern-4mu}

\newcommand{\Nc}{\mathcal{N}}

\newcommand{\Ic}{\mathcal{I}}
\newcommand{\tr}{\mathrm{tr}}
\newcommand{\Ncc}{N_{\mathrm{C}}}

\newcommand{\vx}{\vec{x}}
\newcommand{\vy}{\vec{y}}
\newcommand{\vz}{\vec{z}}
\newcommand{\vp}{\vec{p}}
\newcommand{\vq}{\vec{q}}
\newcommand{\vk}{\vec{k}}
\newcommand{\vA}{\vec{A}}
\newcommand{\vB}{\vec{B}}
\newcommand{\vD}{\vec{D}}
\newcommand{\vPi}{\mathbf{\Pi}}
\newcommand{\valpha}{\boldsymbol{\alpha}}
\newcommand{\hp}{\hat{\vp}}
\newcommand{\hq}{\hat{\vq}}
\newcommand{\hk}{\hat{\vk}}

\newcommand{\hpq}{\widehat{(\vp + \vq)}}

\usepackage[font=small]{caption}
\begin{document}

\title{Revised variational approach to QCD in Coulomb gauge}

\author{Davide R.~Campagnari, Ehsan Ebadati, Hugo Reinhardt and Peter Vastag}
\affiliation{Institut f\"ur Theoretische Physik\\
Eberhard-Karls-Universit\"at T\"ubingen \\
Auf der Morgenstelle 14\\
D-72076 T\"ubingen\\
Germany}
\date{\today}

\begin{abstract}
The variational approach to QCD in Coulomb gauge is revisited. By assuming the non-Abelian Coulomb potential to be given by the sum of its infrared and ultraviolet parts, i.e.~by a linearly rising potential and an ordinary Coulomb potential, and by using a Slater determinant ansatz for the quark wave functional, which contains the coupling of the quarks and the gluons with two different Dirac structures, we obtain variational equations for the kernels of the fermionic vacuum wave functional, which are free of ultraviolet divergences. Thereby, a Gaussian type wave functional is assumed for the gluonic part of the vacuum. By using the results of the pure Yang--Mills sector for the gluon propagator as input, we solve the equations for the fermionic kernels numerically and calculate the quark condensate and the effective quark mass in leading order. Assuming a value of $\sigma_{\mathrm{C}} = 2.5 \sigma$ for the Coulomb string tension (where $\sigma$ is the usual Wilsonian string tension) the phenomenological value of the quark condensate  $\langle \bar{\psi} \psi \rangle \simeq (-235 \, \mathrm{MeV})^3$ is reproduced with a value of $g \simeq 2.1$ for the strong coupling constant of the quark-gluon vertex.
\end{abstract}
\maketitle

\section{Introduction}

In recent years the vacuum sector of Yang--Mills theory was treated within the Hamiltonian approach in Coulomb gauge using the variational principle \cite{Schuette1984, SS2001, Feuchter2004, *Feuchter2004a}. In the approach of ref.~\cite{Feuchter2004, *Feuchter2004a}, the Gribov--Zwanziger confinement scenario \cite{Gribov1978, Zwanziger1998} was found to be realized \cite{ERS2007}: A linearly rising static quark potential, as well as infrared (IR) diverging ghost form factor and gluon energy (see eq.~(\ref{Gl: Gribov}) below) were found. The latter could be nicely fitted by Gribov's formula \cite{Gribov1978, BQR2009}.

In ref.~\cite{Pak2013, *Pak2012a}, the variational approach to Yang--Mills theory in Coulomb gauge was extended to full Quantum Chromodynamics (QCD). Thereby the coupling of the quarks to the gluons was included in the (fermionic) vacuum wave functional by a single Dirac structure corresponding to the quark-gluon coupling in the QCD Hamiltonian. In ref.~\cite{QCDT0}, a second Dirac structure for the quark-gluon coupling was included. Thereby it was observed that the leading (linear) order ultraviolet (UV) divergences cancel in the gap equation for the scalar variational kernel. However, in ref.~\cite{QCDT0} the Coulombic part of the non-Abelian Coulomb term was not properly included due to a sign error. Here we show that with the proper inclusion of the Coulombic part of the non-Abelian Coulomb potential all UV divergences cancel in the quark gap equation. The latter is solved numerically and results are presented for the 
quark condensate and the effective quark mass.

The organization of the paper is as follows: In the next section, we present the QCD Hamiltonian in Coulomb gauge and summarize some results obtained for the pure Yang-Mills theory, which serve as input for the quark sector. The variational ansatz for the QCD vacuum wave functional as well as the variational equations of motion are presented in sec.~\ref{Abschn: Variationsansatz}. The UV behavior of these equations is discussed in sec.~\ref{Abschn: UV}, while the static quark propagator and the chiral condensate are given in sec.~\ref{Abschn: Quarkpropagator}. The numerical solution of the variational equations of motion is presented in sec.~\ref{Abschn: Numerik} and some concluding remarks are given in sec.~\ref{Abschn: Zusammenfassung}.

\section{The QCD-Hamiltonian in Coulomb gauge} \label{Abschn: QCDHamiltonian}

The QCD Hamiltonian in Coulomb gauge, $\nabla \cdot \vA = 0$, reads \cite{Pak2012a}
\beq
H_{\mathrm{QCD}} = H_{\mathrm{YM}} + H_{\mathrm{Q}} + H_{\mathrm{C}} \label{Gl: QCDHamiltonian}
\eeq
where
\beq
H_{\mathrm{YM}} = \frac{1}{2} \int \dd^3 x \left(J^{-1}[A] \, \vPi(\vx) J[A] \, \vPi(\vx) + \vB^2(\vx) \right) \label{Gl: YMHamiltonian}
\eeq
is the Hamiltonian of the transversal components of the gauge field. Here
\beq
\Pi^a_k(\vx) = \frac{\delta}{\ii \delta A^a_k(\vx)} \label{Gl: KanImpuls}
\eeq
is the operator of the canonical momentum of the gluons (which represents the color electric field) and
\beq
B^a_k(\vx) = \varepsilon_{klm} \left(\partial_l A^a_m(\vx) - \frac{g}{2} f^{abc} A^b_l(\vx) A^c_m(\vx)\right) \label{Gl: Farbmagnetfeld}
\eeq
is 
the non-Abelian color magnetic field ($g$ is the bare strong coupling constant and $f$ is the structure constant of the color group). Furthermore,
\beq
J[A] = \det\bigl({\hat{G}}^{-1}\bigr) \label{Gl: FaddeevPopov}
\eeq
is the Faddeev--Popov determinant where
\beq
\bigl(\hat{G}^{-1}\bigr)^{a b}(\vx, \vy) = \bigl(-\nabla \cdot \hat{\vD}\bigr)^{a b}(\vx, \vy)
\eeq
denotes the Faddeev--Popov operator containing the covariant derivative in adjoint representation
\beq
\hat{D}^{a b}_k(\vx) = \delta^{ab} \partial_k - g f^{acb} A^c_k(\vx) \, . \label{Gl: KovAbleitung}
\eeq
The second term in eq.~(\ref{Gl: QCDHamiltonian}) denotes the Dirac Hamiltonian of the quark field $\psi$ interacting with the gauge field $\vA$,\footnote{For simplicity, we consider only one single chiral, i.e.~massless, quark flavor.}
\beq
H_{\mathrm{Q}} = \int \dd^3 x \, \psi^\dagger(\vx) \vec{\alpha} \cdot \bigl(-\ii \nabla + g t^a \vA^a(\vx)\bigr) \psi(\vx) \label{Gl: DiracHamiltonian}
\eeq
with $t$ being the generator of the color group in the fundamental representation. Finally, the third term in eq.~(\ref{Gl: QCDHamiltonian}), the so-called Coulomb term
\beq
H_{\mathrm{C}} = \frac{g^2}{2} \int \dd^3 x \int \dd^3 y \, J^{- 1}[A] \rho^a(\vx) J[A] \hat{F}^{ab}(\vx, \vy) \rho^b(\vy) \, , \label{Gl: Coulombterm}
\eeq
stems from the longitudinal components of the canonical momentum operator after resolving Gau{\ss}'s law. Here
\beq
\hat{F}^{a b}(\vx, \vy) = \int \dd^3 z \, \hat{G}^{a c}(\vx, \vz) (-\Delta_{\vz}) \hat{G}^{c b}(\vz, \vy) \label{Gl: Coulombkern}
\eeq
is the Coulomb kernel and
\beq
\rho^a(\vx) = \rho_{\mathrm{YM}}^a(\vx) + \rho_{\mathrm{Q}}^a(\vx) = f^{abc} \vA^b(\vx) \cdot \vPi^c(\vx) + \psi^\dagger(\vx) t^a \psi(\vx)
\eeq
is the color density of the gluons and quarks. Up to two-loop order in the energy, it is sufficient to replace the Coulomb kernel by its gluonic expectation value
\beq
g^2 \langle \hat{F}^{a b}(\vx, \vy) \rangle_{\mathrm{YM}} = \delta^{a b} V_{\mathrm{C}}(|\vx - \vy|) \label{Gl: Coulombkern3}
\eeq
(which yields the static color potential $V_{\mathrm{C}}$) and to use the Gaussian functional
\beq
J[A] = \exp\left(-\int \dd^3 x \int \dd^3 y \, A_k^a(\vx) \chi_{k l}^{a b}(\vx, \vy) A_l^b(\vy)\right) \label{Gl: FaddeevPopov1}
\eeq
for the Faddeev--Popov determinant where
\beq
\chi_{k l}^{a b}(\vx, \vy) = -\frac{1}{2} \Bigl\langle \frac{\delta}{\delta A_k^a(\vx)} \frac{\delta}{\delta A_l^b(\vy)} \ln J[A] \Bigr\rangle_{\mathrm{YM}} \label{Gl: Kruemmung}
\eeq
is the ghost loop referred to as curvature \cite{Feuchter2005}. The actual calculation performed in ref.~\cite{ERS2007} shows that the Coulomb potential (\ref{Gl: Coulombkern3}) can be nicely fitted by a superposition of a linearly rising and an ordinary Coulomb term, i.e.~by a sum of its IR and UV limits,
\beq
V_{\mathrm{C}}(r) = -\sigma_{\mathrm{C}} r + \frac{\alpha_{\mathrm{S}}}{r} \label{Gl: Coulombkern4}
\eeq
where $\sigma_{\mathrm{C}}$ is the so-called Coulomb string tension and $\alpha_{\mathrm{S}} = g^2/4 \pi$.

\section{Variational approach}

\subsection{Variational ansatz and equations of motion} \label{Abschn: Variationsansatz}

Following ref.~\cite{QCDT0}, we use the ansatz
\beq
\vert \phi[A] \rangle = \phi_{\mathrm{YM}}[A] \, \vert \phi_{\mathrm{Q}}[A] \rangle \label{Gl: Vakuumfunktional}
\eeq
for the QCD vacuum wave functional where the gluonic part is given by the Gaussian type functional
\begin{subequations}
\begin{align}
\phi_{\mathrm{YM}}[A] &=  \Nc I^{-\frac{1}{2}}[A] J^{-\frac{1}{2}}[A] \widetilde{\phi}_{\mathrm{YM}}[A] \, , \\
\widetilde{\phi}_{\mathrm{YM}}[A] &= \exp\left(-\frac{1}{2} \int \dd^3 x \int \dd^3 y \, A_k^a(\vx) \omega(\vx, \vy) A_k^a(\vy)\right) \, ,
\end{align}
\label{Gl: BoseAnsatz1}%
\end{subequations}
with a normalization factor $\Nc$, the fermionic determinant $I = \langle \phi_{\mathrm{Q}} \vert \phi_{\mathrm{Q}} \rangle$ and $\omega$ being a 
variational kernel. For the quark part the ansatz \cite{QCDT0}
\beq
\vert \phi_{\mathrm{Q}}[A] \rangle = \exp\left[-\int \dd^3 x \int \dd^3 y \, \psi_+^{\dagger}(\vx) K(\vx, \vy) \psi_-(\vy)\right] \vert
0 \rangle \, , \label{Gl: VakuumfunktionalQ}
\eeq
is assumed, where $\psi_{\pm}$ denotes the positive/negative spectral projection of the field operator, $\vert 0 \rangle$ is the bare fermionic vacuum (Dirac sea) and
\beq
K(\vx, \vy) = \beta S(\vx, \vy) + g \int \dd^3 z \, \bigl[V(\vx, \vy; \vz) + \beta W(\vx, \vy; \vz)\bigr] \valpha \cdot \vA^a(\vz) t^a
\label{Gl: VakuumfunktionalQ1}
\eeq
contains three variational kernels $S$, $V$, $W$, which, together with $\omega$, have to be determined by minimizing the ground state energy. The ansatz (\ref{Gl: VakuumfunktionalQ}), (\ref{Gl: VakuumfunktionalQ1}) for the quark wave functional reduces for $V = W = 0$ to the BCS-type wave functional used in refs.~\cite{FM1982, Adler1984, LeYaouanc1984, AA1988} and for $W = 0$ to  the ansatz considered in ref.~\cite{Pak2013}.

In ref.~\cite{QCDT0}, the vacuum energy $\langle H_{\mathrm{QCD}} \rangle \equiv \langle \phi \vert H_{\mathrm{QCD}} \vert \phi \rangle$ was calculated with the wave functional (\ref{Gl: Vakuumfunktional}) up to including two loops. This is conveniently done in momentum space. We use the same convention as in ref.~\cite{QCDT0} for the kernels ($\da \equiv \dd / 2 \pi$)
\begin{align}
S(\vx, \vy) &= \int \da^3 p \, \exp\bigl(\ii \vp \cdot (\vx - \vy)\bigr) S(p) \, , \\
V(\vx, \vy; \vz) &= \int \da^3 p \int \da^3 q \, \exp\bigl(\ii \vp \cdot (\vx - \vz)\bigr) \exp\bigl(\ii \vq \cdot (\vy - \vz)\bigr) V(\vp, \vq) 
\end{align}
and analogous definitions for the Fourier transforms of $\omega (\vx, \vy)$ and $W (\vx, \vy; \vz)$.  Here we have exploited translational and rotational invariance and overall momentum conservation. The quark field is expanded as
\begin{align}
\psi^m(\vx) &= \int \da^3 p \, \frac{1}{\sqrt{2 p}} \exp(\ii \vp \cdot \vx) \Bigl(a^{s, m}(\vp) u^s(\vp) + {b^{s, m}}^{\dagger}(-\vp) v^s(-\vp)\Bigr)
\end{align}
where $a$ ($b$) denotes the annihilation operator for a (anti-)quark state and $u$ ($v$) is the Dirac eigenspinor 
with positive (negative) eigenvalue. Furthermore, $s = \pm 1$ is the double of the spin projection.

Variation of $\langle H_{\mathrm{QCD}} \rangle$ with respect to the scalar kernel $S$ yields the following integral equation \cite{QCDT0}
\beq
k S(k) = I_{\mathrm{C}}^{\mathrm{Q}}(k) + I_{V V}^{\mathrm{Q}}(k) + I_{W W}^{\mathrm{Q}}(k) + I_{V \mathrm{Q}}^{\mathrm{Q}}(k) + I_{W \mathrm{Q}}^{\mathrm{Q}}(k) + I_{E}^{\mathrm{Q}}(k) \label{Gl: Gapgleichung}
\eeq
to which we will refer as (quark) gap equation. Here,
\beq
I_{\mathrm{C}}^{\mathrm{Q}}(k) = \frac{C_{\mathrm{F}}}{2} \int \da^3 p \, V_{\mathrm{C}}(|\vp - \vk|) P(p) \left[S(p) \bigl(1 - S^2(k)\bigr) - S(k) \bigl(1 - S^2(p)\bigr) \hp \cdot \hk\right] \label{Gl: CoulombQQGapgl}
\eeq
is the contribution of the Coulomb term $H_{\mathrm{C}}$ (\ref{Gl: Coulombterm}) with the Casimir factor $C_{\mathrm{F}} = (\Ncc^2 - 1) / 2 \Ncc$ and
\beq
V_{\mathrm{C}}(p) = \frac{8 \pi \sigma_{\mathrm{C}}}{p^4} + \frac{4 \pi \alpha_{\mathrm{S}}}{p^2} = V_{\mathrm{C}}^{\mathrm{IR}}(p) + V_{\mathrm{C}}^{\mathrm{UV}}(p) \label{Gl: Coulombkern2}
\eeq
being the Coulomb potential (\ref{Gl: Coulombkern4}) in momentum space. Furthermore, 
\begin{align}
I_{V V}^{\mathrm{Q}}(k) &= -\frac{C_{\mathrm{F}}}{2} g^2 \int \da^3 p \, \frac{V^2(\vp, \vk)}{\omega(|\vp + \vk|)} X(\vp, \vk) P(p) \Bigl\{k P(k) S(k)\Bigl[-3 + S^2(k)\Bigr] + p P(p) S(k) \Bigl[-1 + S^2(p)\Bigr] \nonumber \\
&\phantom{=}\,\, \phantom{-\frac{C_{\mathrm{F}}}{2} g^2 \int \da^3 p \, \times \Bigl\{} + k P(k) S(p) \Bigl[1 - 3 S^2(k)\Bigr] + p P(p) S(p) \Bigl[1 - S^2(k)\Bigr]\Bigr\} \, , \label{Gl: FrDiracGapglV} \\
I_{W W}^{\mathrm{Q}}(k) &= - \frac{C_{\mathrm{F}}}{2} g^2 \int \da^3 p \, \frac{W^2(\vp, \vk)}{\omega(|\vp + \vk|)} Y(\vp, \vk) P(p) \Bigl\{k P(k) S(k) \Bigl[-3 + S^2(k)\Bigr] + p P(p) S(k) \Bigl[-1 + S^2(p)\Bigr] \nonumber \\
&\phantom{=}\,\, \phantom{- \frac{C_{\mathrm{F}}}{2} g^2 \int \da^3 p \, \times \Bigl\{} - k P(k) S(p) \Bigl[1 - 3 S^2(k)\Bigr] - p P(p) S(p) \Bigl[1 - S^2(k)\Bigr]\Bigr\} \label{Gl: FrDiracGapglW}
\end{align}
result from the free single particle Dirac Hamiltonian,
\begin{align}
I_{V \mathrm{Q}}^{\mathrm{Q}}(k) &= \frac{C_{\mathrm{F}}}{2} g^2 \int \da^3 p \, \frac{V(\vp, \vk)}{\omega(|\vp + \vk|)} X(\vp, \vk) P(p) \Bigl[S(p) \bigl(1 - S^2(k)\bigr) - 2 S(k)\Bigr] \label{Gl: KopplungstermGapglV} \, , \\
I_{W \mathrm{Q}}^{\mathrm{Q}}(k) &= \frac{C_{\mathrm{F}}}{2} g^2 \int \da^3 p \, \frac{W(\vp, \vk)}{\omega(|\vp + \vk|)} Y(\vp, \vk) P(p) \Bigl[1 - S^2(k) - 2 S(k) S(p)\Bigr] \label{Gl: KopplungstermGapglW}
\end{align}
are the contributions stemming from the quark-gluon coupling in the Dirac Hamiltonian $H_{\mathrm{Q}}$ (\ref{Gl: DiracHamiltonian}) and, finally,
\beq
I_{E}^{\mathrm{Q}}(k) = \frac{C_{\mathrm{F}}}{2} g^2 S(k) \int \da^3 p \, V^2(\vp, \vk) X(\vp, \vk) P(p) + \frac{C_{\mathrm{F}}}{2} g^2 S(k) \int \da^3 p \, W^2(\vp, \vk) Y(\vp, \vk) P(p) \label{Gl: KinetischeEnergieGapgl}
\eeq
results from the action of the operator of the gluonic kinetic energy $H_{\mathrm{YM}}$ (\ref{Gl: YMHamiltonian}) on the quark wave functional. In the above equations, we have used the abbreviations ($\hp = \vp / p$)
\begin{align}
P(p) &= \frac{1}{1 + S^2(p)} \, , \label{Gl: PFaktor} \\
X(\vp, \vq) &= 1 - \Bigl[\hp \cdot \hpq\Bigr] \Bigl[\hq \cdot \hpq\Bigr] \, ,  \\
Y(\vp, \vq) &= 1 + \Bigl[\hp \cdot \hpq\Bigr] \Bigl[\hq \cdot \hpq\Bigr] \, .
\end{align}
The variational equations for the quark-gluon coupling kernels $V$ and $W$ can be explicitly solved in terms of $S(p)$ and $\omega(p)$ yielding \cite{QCDT0}
\beq
V(\vk, \vk') = \frac{1 + S(k) S(k')}{k P(k) \Bigl(1 - S^2(k) + 2 S(k) S(k')\Bigr) + k' P(k') \Bigl(1 - S^2(k') + 2 S(k) S(k')\Bigr) + \omega(|\vk + \vk'|)} \label{Gl: VKern}
\eeq
and
\beq
W(\vk, \vk') = \frac{S(k) + S(k')}{k P(k) \Bigl(1 - S^2(k) - 2 S(k) S(k')\Bigr) + k' P(k') \Bigl(1 - S^2(k') - 2 S(k) S(k')\Bigr) + \omega(|\vk + \vk'|)} \, . \label{Gl: WKern}
\eeq
In principle, our approach yields also a variational integral equation for the gluon propagator $\sim \omega^{- 1} (p)$, see ref.~\cite{QCDT0}. However, here we perform a quenched calculation and use for $\omega(p)$ Gribov's formula \cite{Gribov1978}
\beq
\omega(p) = \sqrt{p^2 + \frac{M_{\mathrm{G}}^4}{p^2}} \label{Gl: Gribov}
\eeq
which nicely fits the lattice data with a Gribov mass of $M_{\mathrm{G}} \simeq 880 \, \mathrm{MeV}$ \cite{BQR2009}.

\subsection{UV-behavior} \label{Abschn: UV}

Assuming that the scalar kernel $S$ is vanishing sufficiently fast in the UV as expected from asymptotic freedom, one finds that the loop terms on the r.h.s.~of the gap equation (\ref{Gl: Gapgleichung}) containing the vector kernel $V$ yield the UV divergence
\beq
\frac{C_{\mathrm{F}}}{16 \pi^2} g^2 S(k) \left[-2 \Lambda + k \ln \frac{\Lambda}{\mu} \left(-\frac{2}{3} + 4 P(k)\right)\right] \label{254-33}
\eeq
($\Lambda$ is the UV cutoff and $\mu$ an arbitrary momentum scale) while the loop terms containing the vector kernel $W$ give
\beq
\frac{C_{\mathrm{F}}}{16 \pi^2} g^2 S(k) \left[2 \Lambda + k \ln \frac{\Lambda}{\mu} \left(\frac{10}{3} - 4 P(k)\right)\right] \, .
\label{270-34}
\eeq
Finally, the loop contribution (\ref{Gl: CoulombQQGapgl}) of the Coulomb potential gives rise to the UV divergence\footnote{Note that this UV divergence is exclusively stemming from the UV part of the Coulomb potential $V_{\mathrm{C}}^{\mathrm{UV}}(p)$ (\ref{Gl: Coulombkern2}) while its IR 
part $V_{\mathrm{C}}^{\mathrm{IR}}(p)$ yields UV finite contributions.}
\beq
\label{277-35}
-\frac{C_{\mathrm{F}}}{6 \pi^2} g^2 k S(k) \ln \frac{\Lambda}{\mu} \, .
\eeq
The crucial point now is that the sum of these UV divergent contributions vanish so that the quark gap equation (\ref{Gl: Gapgleichung}) is in fact UV finite. As one observes from eqs.~(\ref{254-33}) and (\ref{270-34}), the cancellation of the linear UV divergences requires the inclusion of both Dirac structures of the quark-gluon coupling in the vacuum wave functional (\ref{Gl: VakuumfunktionalQ}), (\ref{Gl: VakuumfunktionalQ1}). Cancellation of the logarithmic UV divergences demands in addition the inclusion of the UV part of the Coulomb potential, $V_{\mathrm{C}}^{\mathrm{UV}}(p)$ (\ref{Gl: Coulombkern2}).\footnote{In ref.~\cite{QCDT0} due to the wrong sign of the Coulombic term $V_{\mathrm{C}}^{\mathrm{UV}}(p)$ (\ref{Gl: Coulombkern2}) the cancellation of the logarithmic UV divergences was missed.}

\subsection{Static quark propagator and chiral condensate} \label{Abschn: Quarkpropagator}

The static quark propagator
\beq
G_{i j}^{m n}(\vx, \vy) = \frac{1}{2} \bigl\langle \Bigl[\psi_i^m(\vx), {\psi_j^n}^{\dagger}(\vy)\Bigr] \bigr\rangle \label{Gl: StatProp}
\eeq
can be calculated along the same lines as the ground state energy and reads in momentum space (up to including one-loop terms) \cite{QCDT0}
\beq
G(\vp) = \frac{P(p)}{2} \Bigl[1 - S^2(p) - I_{\alpha}(p)\Bigr] \valpha \cdot \hp + P(p) \Bigl[S(p) - I_{\beta}(p)\Bigr] \beta 
\label{Gl: StatProp1}
\eeq
where the loop terms are given by
\begin{align}
I_{\alpha}(p) &= C_{\mathrm{F}} g^2 \int \da^3 q \, \frac{P(p) P(q)}{\omega(|\vp + \vq|)} \left[V^2(\vp, \vq) X(\vp, \vq) \Bigl(1 + 2 S(p) S(q) - S^2(p)\Bigr) \right. \nonumber \\
&\phantom{=}\,\, \phantom{C_{\mathrm{F}} g^2 \int \da^3 q \, \frac{P(p) P(q)}{\omega(|\vp + \vq|)} \left[\right.} \left. + W^2(\vp,
\vq) Y(\vp, \vq) \Bigl(1 - 2 S(p) S(q) - S^2(p)\Bigr)\right], \label{Gl: Schleifenintegralalpha} \\
I_{\beta}(p) &= \frac{C_{\mathrm{F}}}{2} g^2 \int \da^3 q \, \frac{P(p) P(q)}{\omega(|\vp + \vq|)} \left[V^2(\vp, \vq) X(\vp, \vq) \Bigl(2 S(p) - S(q) + S^2(p) S(q)\Bigr) \right. \nonumber \\
&\phantom{=}\,\, \phantom{\frac{C_{\mathrm{F}}}{2} g^2 \int \da^3 q \, \frac{P(p) P(q)}{\omega(|\vp + \vq|)} \left[\right.} \left. + 
W^2(\vp, \vq) Y(\vp, \vq) \Bigl(2 S(p) + S(q) - S^2(p) S(q)\Bigr)\right] \, . \label{Gl: Schleifenintegralbeta}
\end{align}
The UV analysis of these loop contributions yields the following, divergent behavior:
\begin{align}
I_{\alpha}(p) &= \frac{C_{\mathrm{F}} g^2}{8 \pi^2} \Bigl(1 - S^2(p)\Bigr) \ln \frac{\Lambda}{\mu} + \text{finite terms} \label{Gl: StatPropDiv1} \\
I_{\beta}(p) &= \frac{C_{\mathrm{F}} g^2}{8 \pi^2} S(p) \ln \frac{\Lambda}{\mu} + \text{finite terms} \,. \label{Gl: StatPropDiv2}
\end{align}

The quark propagator (\ref{Gl: StatProp1}) can be rewritten in the quasi-particle form
\beq
G(\vp) = \widetilde{Z}(p) \frac{\valpha \cdot \vp + \beta \widetilde{M}(p)}{2 \sqrt{p^2 + \widetilde{M}^2(p)}} \label{Gl: StatProp3}
\eeq
with an effective quark mass function
\beq
\widetilde{M}(p) = \frac{2 p \bigl[S(p) - I_{\beta}(p)\bigr]}{1 - S^2(p) - I_{\alpha}(p)} \label{Gl: Massenfkt}
\eeq
and the field renormalization factor
\beq
\widetilde{Z}(p) = P(p) \sqrt{\bigl[1 - S^2(p) - I_{\alpha}(p)\bigr]^2 + 4 \bigl[S(p) - I_{\beta}(p)\bigr]^2} \,. \label{Gl: Feldrenormierung}
\eeq
From the expression (\ref{Gl: StatProp3}) for the static quark propagator one finds for the chiral quark condensate
\beq
\langle \bar{\psi}(\vx) \psi(\vx) \rangle = - \tr\bigl(\beta G(\vx, \vx)\bigr) = -2 \Ncc \int \da^3 p \, \frac{\widetilde{Z}(p) \widetilde{M}(p)}{\sqrt{p^2 + \widetilde{M}^2(p)}} \, . \label{Gl: ChiralesKondensat4}
\eeq
Spontaneous breaking of chiral symmetry, $\langle \bar{\psi} \psi \rangle \neq 0$, obviously requires a non-vanishing mass function $\widetilde{M}$ (\ref{Gl: Massenfkt}) (or scalar kernel $S$).  

The one-loop terms (\ref{Gl: Schleifenintegralalpha}), (\ref{Gl: Schleifenintegralbeta}) in the  propagator (\ref{Gl: StatProp1}) give rise to two-loop terms in the quark condensate. When their UV-divergent pieces are removed by counterterms in a minimal subtraction scheme, we find that the finite contributions to the loop integrals (\ref{Gl: Schleifenintegralalpha}), (\ref{Gl: Schleifenintegralbeta}) have only small effect (some percent) to the quark condensate and will hence be ignored in the following. The quark condensate is then given by
\beq
\langle \bar{\psi}(\vx) \psi(\vx) \rangle = -2 \Ncc \int \da^3 p \, \frac{M(p)}{E(p)}
\eeq
where the mass function (\ref{Gl: Massenfkt}) is now given by
\beq
M(p) = \frac{2 p S(p)}{1 - S^2(p)} \label{350-46}
\eeq
and
\beq
E(p) = \sqrt{p^2 + M^2(p)}
\eeq
plays the role of a quasi-particle energy.

\section{Numerical results} \label{Abschn: Numerik}

For the numerical solution of the quark gap equation (\ref{Gl: Gapgleichung}) it is convenient to rewrite it in terms of the mass function (\ref{350-46}). This yields
\beq
\label{346-9}
M(k) = \Ic_{\mathrm{C}}^{\mathrm{Q}}(k) + \Ic_{V V}^{\mathrm{Q}}(k) + \Ic_{W W}^{\mathrm{Q}}(k) + \Ic_{V \mathrm{Q}}^{\mathrm{Q}}(k) + 
\Ic_{W \mathrm{Q}}^{\mathrm{Q}}(k) + \Ic_{E}^{\mathrm{Q}}(k) 
\eeq
where the loop terms on the r.h.s.~are given by
\begin{align}
\Ic_{\mathrm{C}}^{\mathrm{Q}}(k) &= \frac{C_{\mathrm{F}}}{2} \int \da^3 p \, V_{\mathrm{C}}(|\vp + \vk|) \frac{M(p) + M(k) \frac{\vp \cdot \vk}{k^2}}{E(p)} \, , \\
\Ic_{V V}^{\mathrm{Q}}(k) &= -\frac{C_{\mathrm{F}}}{2} g^2 \int \da^3 p \, \frac{V^2(\vp, \vk)}{\omega(|\vp + \vk|)} X(\vp, \vk) \left\{-\frac{E(p) + p}{2 E(p)} M(k) \frac{E(k) + 2 k}{E(k)} \right. \nonumber \\
&\phantom{=}\,\, \phantom{-\frac{C_{\mathrm{F}}}{2} g^2 \int \da^3 p} \left. - p^2 \frac{E(p) + p}{2 E^2(p)} \frac{M(k)}{k} + \frac{M(p)}{2 E(p)} \frac{E(k) + k}{E(k)} \bigl[-E(k) + 2 k\bigr] + p M(p) \frac{E(p) + p}{2 E^2(p)}\right\} \, , \\
\Ic_{W W}^{\mathrm{Q}}(k) &= -\frac{C_{\mathrm{F}}}{2} g^2 \int \da^3 p \, \frac{W^2(\vp, \vk)}{\omega(|\vp + \vk|)} Y(\vp, \vk) \left\{-\frac{E(p) + p}{2 E(p)} M(k) \frac{E(k) + 2 k}{E(k)} \right. \nonumber \\
&\phantom{=}\,\, \phantom{-\frac{C_{\mathrm{F}}}{2} g^2 \int \da^3 p} \left. - p^2 \frac{E(p) + p}{2 E^2(p)} \frac{M(k)}{k} - \frac{M(p)}{2 E(p)} \frac{E(k) + k}{E(k)} \bigl[-E(k) + 2 k\bigr] - p M(p) \frac{E(p) + p}{2 E^2(p)}\right\} \, , \\
\Ic_{V \mathrm{Q}}^{\mathrm{Q}}(k) &= \frac{C_{\mathrm{F}}}{2} g^2 \int \da^3 p \, \frac{V(\vp, \vk)}{\omega(|\vp + \vk|)} X(\vp, \vk) \left[\frac{M(p)}{E(p)} - \frac{E(p) + p}{E(p)} \frac{M(k)}{k}\right] , \\
\Ic_{W \mathrm{Q}}^{\mathrm{Q}}(k) &= \frac{C_{\mathrm{F}}}{2} g^2 \int \da^3 p \, \frac{W(\vp, \vk)}{\omega(|\vp + \vk|)} Y(\vp, \vk) \left[\frac{E(p) + p}{E(p)} - \frac{M(p)}{E(p)} \frac{M(k)}{k}\right] , \\
\Ic_{E}^{\mathrm{Q}}(k) &= \frac{C_{\mathrm{F}}}{2} g^2 \frac{M(k)}{k} \int \da^3 p \, V^2(\vp, \vk) X(\vp, \vk) \frac{E(p) + p}{2 E(p)} + \frac{C_{\mathrm{F}}}{2} g^2 \frac{M(k)}{k} \int \da^3 p \, W^2(\vp, \vk) Y(\vp, \vk) \frac{E(p) + p}{2 E(p)}
\end{align}
while the vector kernels (\ref{Gl: VKern}), (\ref{Gl: WKern}) read
\beq
V(\vp, \vq) = \frac{1 + \frac{E(p) - p}{M(p)} \frac{E(q) - q}{M(q)}}{\frac{p^2}{E(p)} \left[1 + \frac{M(p)}{p} \frac{E(q) - q}{M(q)}\right] + \frac{q^2}{E(q)} \left[1 + \frac{M(q)}{q} \frac{E(p) - p}{M(p)}\right] + \omega(|\vp + \vq|)}
\eeq
and
\beq
W(\vp, \vq) = \frac{\frac{E(p) - p}{M(p)} + \frac{E(q) - q}{M(q)}}{\frac{p^2}{E(p)} \left[1 - \frac{M(p)}{p} \frac{E(q) - q}{M(q)}\right] + \frac{q^2}{E(q)} \left[1 - \frac{M(q)}{q} \frac{E(p) - p}{M(p)}\right] + \omega(|\vp + \vq|)} \, .
\eeq
Let us stress that the transformation of the gap equation (\ref{Gl: Gapgleichung}) for $S$ to the equation (\ref{346-9}) for $M$ (\ref{350-46}) is exact, i.e.~equations (\ref{Gl: Gapgleichung}) and (\ref{346-9}) are completely equivalent even if eq.~(\ref{350-46}) is only the leading-order expression for the mass function $\widetilde{M}$ (\ref{Gl: Massenfkt}).

In the following, we make some remarks on the numerical solution of the quark gap equation (\ref{346-9}). In the limit $g = 0$, this equation was already solved in a number of previous papers, see ~refs.~\cite{AA1988, Watson2012}. However, the numerical method given e.g.~in ref.~\cite{Watson2012} is not applicable to the full equation (\ref{346-9}). This is because this method separates an IR finite term into two IR divergent terms, which would suppress the remaining IR finite terms of the full equation (\ref{346-9}).

\begin{figure}
\centering
\includegraphics[width=0.4\linewidth]{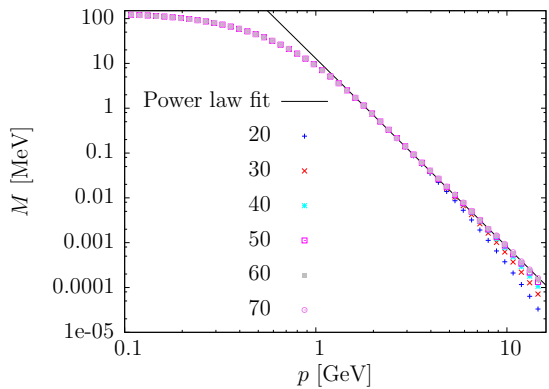}
\caption{Numerical solution of the gap equation (\ref{346-9}) for $g = 0$ for different numbers of $z$-integration points in physical units and on logarithmic scale. The straight line shows a power law fit to the data for 20 integration points and momenta between $1.7 \, \mathrm{GeV} < p < 3.0 \, \mathrm{GeV}$.}
\label{Abb: Winkelintegration}
\end{figure}

In order to solve the gap equation (\ref{346-9}), we first shift the loop momentum $\vp + \vk \to \vq$ which simplifies the handling of the apparent IR divergence of $V_{\mathrm{C}}^{\mathrm{IR}}(q)$ at $\vq = 0$. After switching to spherical coordinates for $\vq$, the integration over the azimuthal angle becomes trivial yielding a factor of $2 \pi$ while for the polar angle the common substitution $\hq \cdot \hk = z$ is used. The remaining integrations over $z$ and $|\vq| = q$ are carried out by means of a standard Gau\ss--Legendre quadrature thereby introducing finite IR ($\kappa$) and UV ($\lambda$) cutoffs for the $q$-integration.\footnote{Note that the IR cutoff serves as regulator for the apparent divergence of the Coulomb term.} The numerical solution is stable for reasonable values of the cutoffs ($\kappa > 0.8 \, \mathrm{MeV}$, $\lambda < 16 \, \mathrm{GeV}$). The number of sampling points for the $z$-integration manifests itself in the numerical result as second scale (beside the physical scale given by the Coulomb string tension $\sigma_{\mathrm{C}}$). This can be clearly seen in fig.~\ref{Abb: Winkelintegration} where the mass function for $g = 0$ is presented on a logarithmic scale. In the UV, the numerical solution shows a power-law behavior up to a critical momentum where a bending sets in. The appearance of this critical momentum is an artifact of our numerical procedure. Increasing the number of integration points of the angular integral  shifts this critical momentum to higher values. For simplicity, we calculate the numerical solution only for a moderate number of sampling points ($\sim 30$) and determine the UV behavior of $M (p)$ by fitting it to a power-law.

In the numerical calculation, we use a Coulomb string tension of $\sigma_{\mathrm{C}} = 2.5 \sigma$, where $\sigma = (440 \, \mathrm{MeV})^2$ is the Wilsonian string tension. This value is favored by the lattice calculation reported in ref.~\cite{BQRV2015}. The quark-gluon coupling constant $g$ is adjusted to reproduce the phenomenological value of the quark condensate $\langle \bar{\psi} \psi \rangle \simeq (-235 \, \mathrm{MeV})^3$ \cite{Williams2007}. This yields $g \simeq 2.1$, which corresponds to a value of the running coupling constant (calculated in ref.~\cite{ERS2007} 
from the ghost-gluon vertex) in the mid-momentum regime.\footnote{The obtained IR value of the running coupling constant is $g = \sqrt{8 \pi^2 / \Ncc} \simeq 5.13$ for $SU(3)$ \cite{Leder2011}.}

\begin{figure}%
\centering%
\parbox{0.4\linewidth}{%
\centering%
\includegraphics[width=\linewidth]{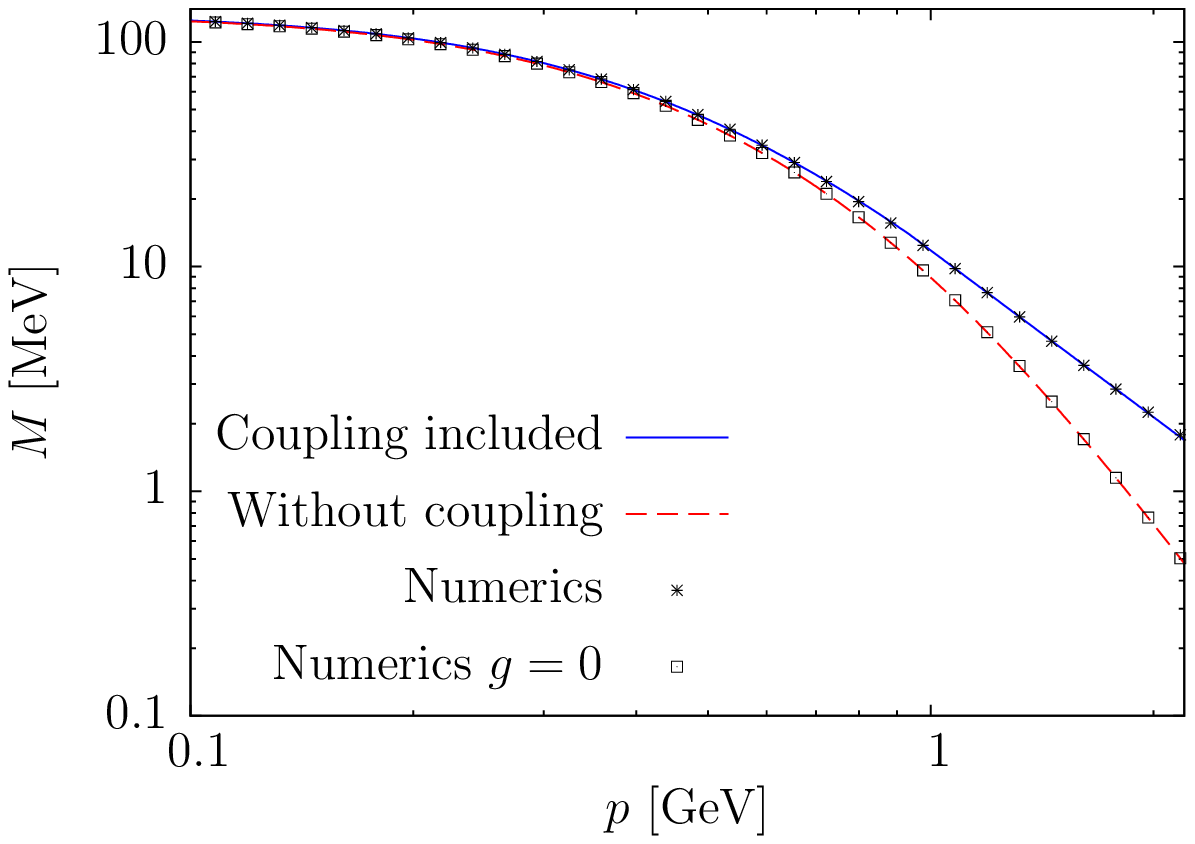} \\%
(a) %
}%
\hspace{0.1\linewidth}%
\parbox{0.4\linewidth}{%
\centering%
\includegraphics[width=\linewidth]{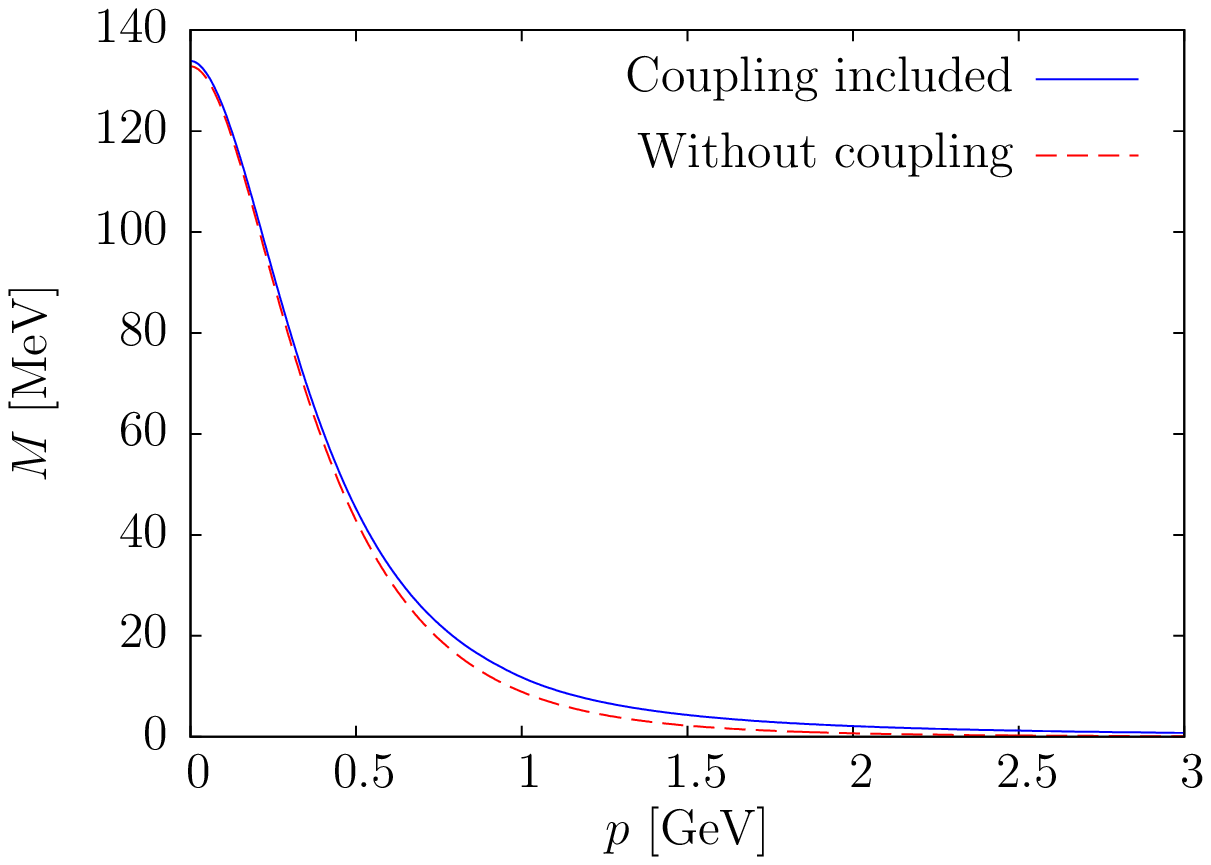} \\%
(b) %
}%
\caption{Numerical solution of the quark gap equation (\ref{346-9}) for the mass function $M$ (\ref{350-46}) comparing the results for $g = 2.1$ (full curve) and $g = 0$ (dashed curve). Differences occur mostly in the UV as can be seen on a logarithmic scale (a) while on a linear scale (b) both solutions show almost the same behavior. Note that the straight lines refer to fitting functions while numerical data points are marked by crosses/boxes.}%
\label{Abb: Massenfunktion}%
\end{figure}%

\begin{figure}%
\centering%
\parbox{0.4\linewidth}{%
\centering%
\includegraphics[width=\linewidth]{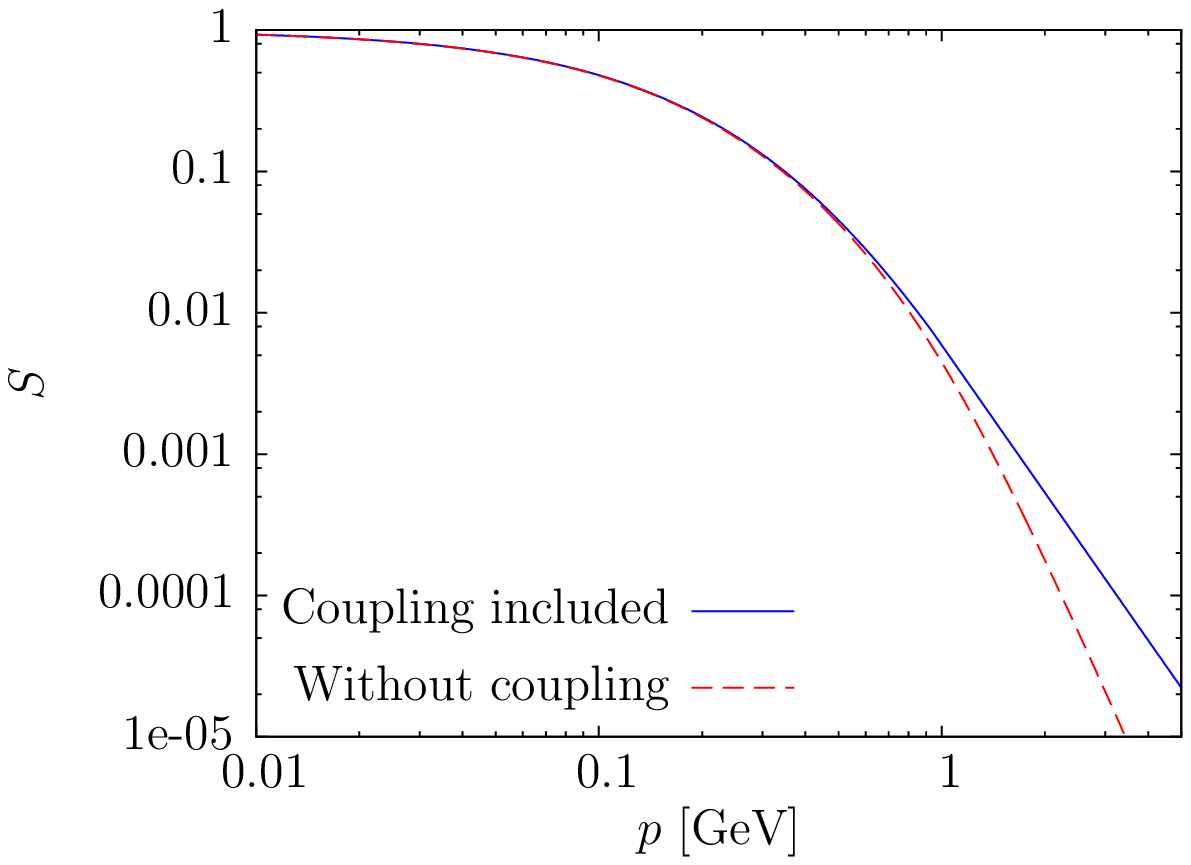} \\%
(a) %
}%
\hspace{0.1\linewidth}%
\parbox{0.4\linewidth}{%
\centering%
\includegraphics[width=\linewidth]{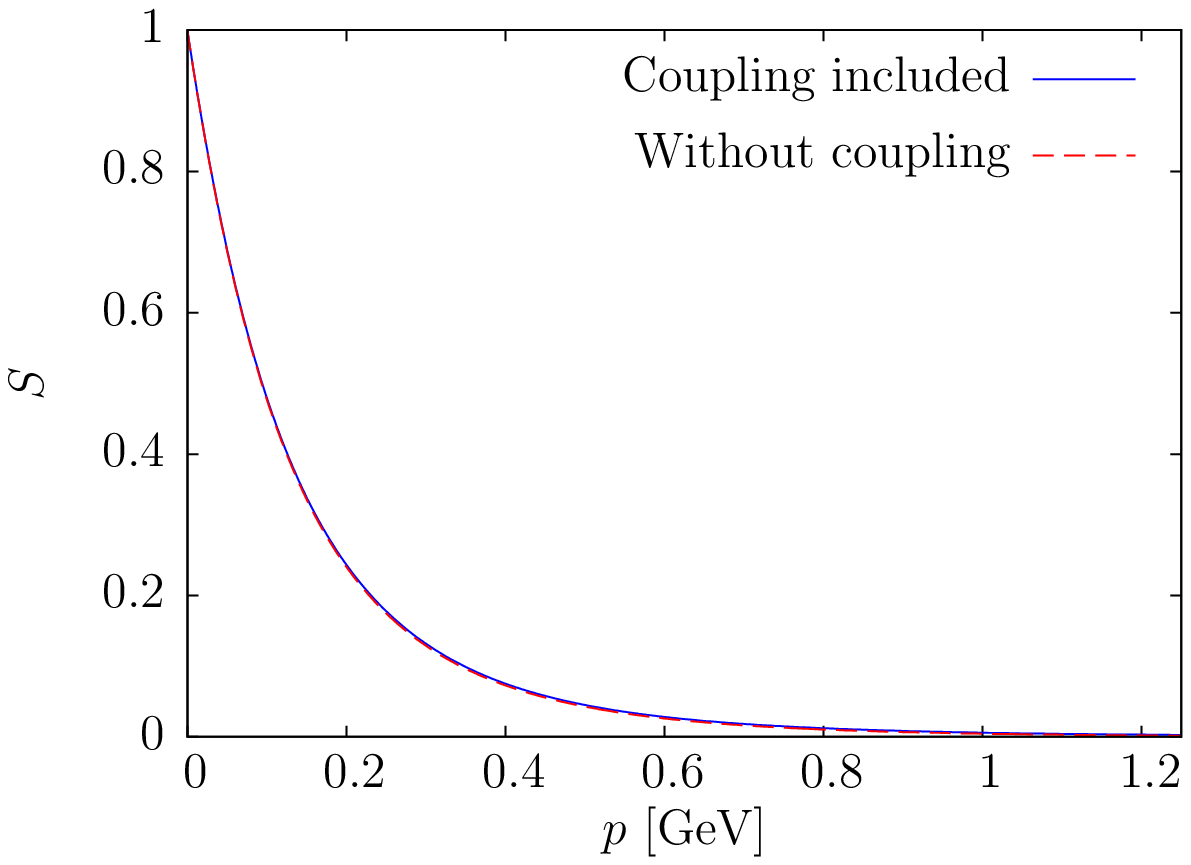} \\%
(b) %
}%
\caption{Numerical solution of the quark gap equation (\ref{346-9}) for the scalar kernel $S$ for $g = 2.1$ (full curve) and $g = 0$ (dashed curve) on a (a) logarithmic and (b) linear scale.}%
\label{Abb: Skalarkern}%
\end{figure}%

Figures \ref{Abb: Massenfunktion} and \ref{Abb: Skalarkern} show the numerical solution of the quark gap equation (\ref{346-9}) for the mass function $M$ (\ref{350-46}) and the scalar kernel $S$, respectively. For sake of comparison we also show the solution when the coupling of the quarks to the transversal gluon is neglected ($g = 0$, Adler--Davis model \cite{Adler1984}). As one observes the inclusion of the coupling to the transversal gluons does not practically alter the IR behavior of $S$ and $M$, while it does change the mid- and large momentum regime. This comes with no surprise: The IR behavior of the gap equation (\ref{Gl: Gapgleichung}), (\ref{346-9}) is dominated by the IR part  of the Coulomb potential, $V_{\mathrm{C}}^{\mathrm{IR}}(p) \sim 1/p^4$, which is present also in the Adler--Davis model. Therefore we expect the same IR behavior for $g = 0$ and $g \neq 0$. The coupling of the quarks to the gluons induces terms in the gap equation (\ref{Gl: Gapgleichung}), which are IR subleading and, in fact, are of the same order as the term arising from the UV part of the Coulomb potential $V_{\mathrm{C}}^{\mathrm{UV}}(p)$ (\ref{Gl: Coulombkern2}), as the cancellation of the UV divergences shows, see eqs.~(\ref{254-33}), (\ref{270-34}) and (\ref{277-35}). If the linearly rising part of the non-Abelian Coulomb potential $V_{\mathrm{C}}^{\mathrm{IR}}$, eq.~(\ref{Gl: Coulombkern4}), is neglected ($\sigma_{\mathrm{C}} = 0$), only the trivial solution is found, $M(p) = 0$, implying that chiral symmetry is not broken spontaneously.

For the calculation of the quark condensate we fit the mass function $M$ for small- and mid-momenta by the analytic expression
\beq
M_{\mathrm{fit}}^{\mathrm{IR}}(p) = \frac{m_0}{1 + \left(\frac{p}{m_A}\right)^A + \left(\frac{p}{m_B}\right)^B} \, . \label{Gl: FitfunktionIR}
\eeq
For $g = 2.1$ the optimized fit parameters read
\begin{alignat}{3}
m_0 &= 134 \, \mathrm{MeV} \, &\quad \quad m_A &= 674 \, \mathrm{MeV} \, & \quad \quad m_B &= 388 \, \mathrm{MeV} \nonumber \\
A &= 3.598 \, &\quad \quad B &= 1.915 \, . & &
\end{alignat}
Above $p \simeq 1 \, \mathrm{GeV}$, we use the power law fit
\beq
M_{\mathrm{fit}}^{\mathrm{UV}}(p) = m_C \left(\frac{p}{m_C}\right)^C \label{Gl: FitfunktionUV}
\eeq
with the fit parameters $m_C = 278 \, \mathrm{MeV}$ and $C = -2.467$. As can be seen from fig.~\ref{Abb: Massenfunktion} (a), this yields a suitable fit to the numerical data points. From eq.~(\ref{Gl: FitfunktionIR}), we can conclude that the IR limit of the mass function is given by $M(p \to 0) \simeq 134 \, \mathrm{MeV}$ which is almost the same as for the Adler--Davis model ($133 \, \mathrm{MeV}$).  However, the UV exponent $C$ obtained from (\ref{Gl: FitfunktionUV}) is much higher than that of the numerical solution for $g = 0$ ($-4.54$).\footnote{Numerical calculations show that both $M(p \to 0)$ and the UV exponent are increasing the higher the coupling $g$ is chosen. However, $M(0)$ only differs significantly from its $g = 0$ value at higher values of the coupling $g > 5$.} The larger UV exponent implies a larger quark condensate. At $g = 2.1$, the chiral condensate obtained reaches its phenomenological value $\langle \bar{\psi} \psi \rangle \simeq (-235 \, \mathrm{MeV})^3$ which is significantly larger than that of the Adler--Davis model, $(-185 \, \mathrm{MeV})^3$.

\begin{figure}%
\centering%
\parbox{0.48\linewidth}{%
\centering%
\includegraphics[width=\linewidth]{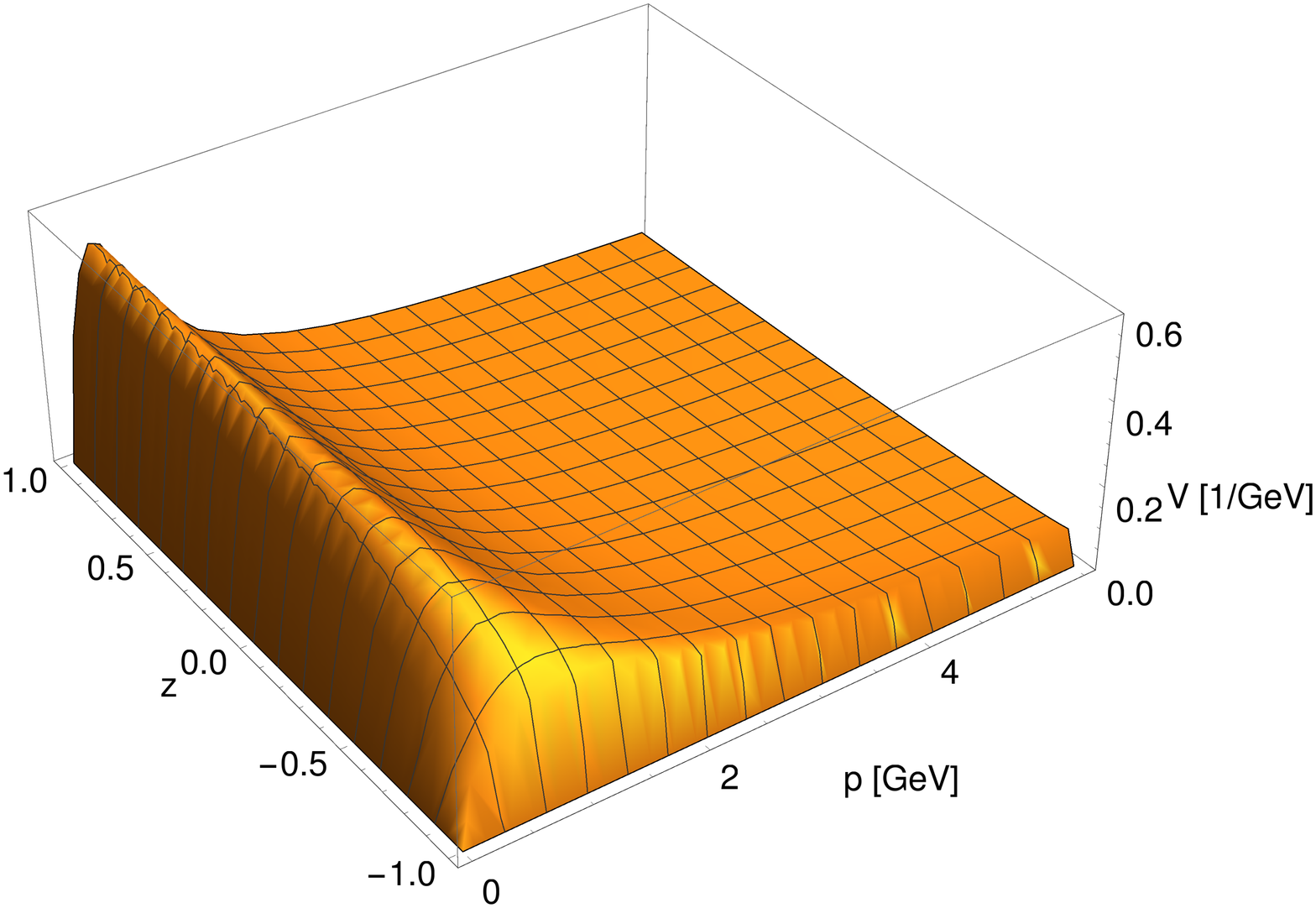} \\%
(a) %
}%
\hfill%
\parbox{0.48\linewidth}{%
\centering%
\includegraphics[width=\linewidth]{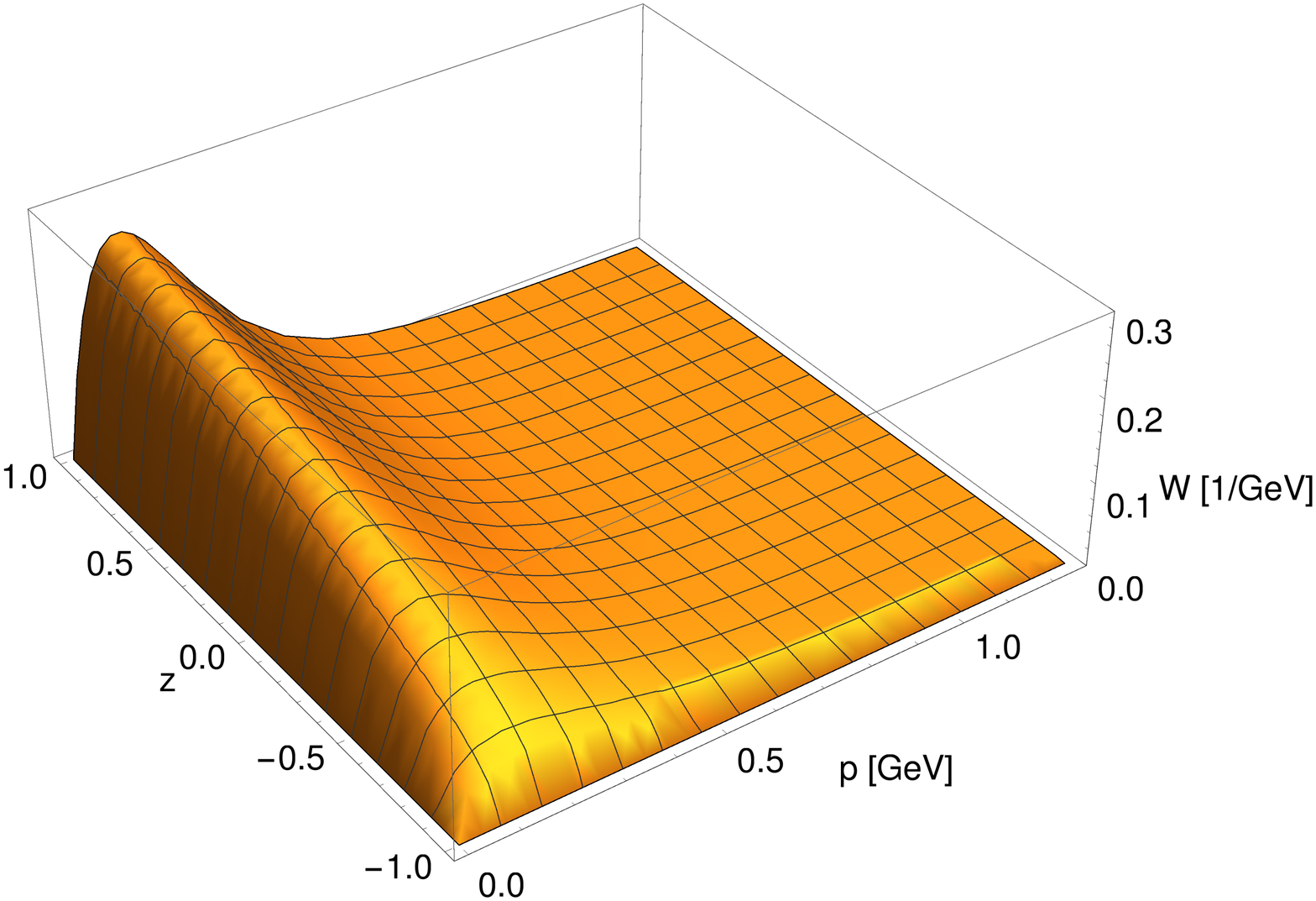} \\%
(b) %
}%
\caption{The vector kernel (a) $V(\vp, \vq)$ and (b) $W(\vp, \vq)$ obtained from the solution of the gap equation (\ref{346-9}) for $g = 2.1$ as a function of the modulus $p = q$ and $z = \cos\sphericalangle(\vp,\vq)$. Note the different scales in subfigure (a) and (b).}%
\label{Abb: Vektorkerne}%
\end{figure}%

\begin{figure}
\centering
\includegraphics[width=0.4\linewidth]{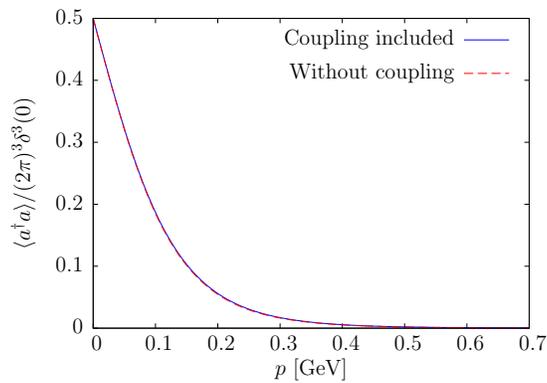}
\caption{Density of occupied quark states for $g = 2.1$ (full curve) and $g = 0$ (dashed curve).}
\label{Abb: Besetzungszahl}
\end{figure}

Using the algebraic fit (\ref{Gl: FitfunktionIR}), (\ref{Gl: FitfunktionUV}), we find for the vector kernels $V(\vp, \vq)$ (\ref{Gl: VKern}) and $W(\vp, \vq)$ (\ref{Gl: WKern}) the result shown in fig.~\ref{Abb: Vektorkerne} for the section $p = q$. Although both kernels have a similar shape, the (non-perturbative) $W$ kernel is significantly smaller than the $V$ kernel. Due to the choice $p = q$, $W$ vanishes much faster in the UV than $V$. However, for a general $q \neq p$, both vector kernels vanish $\sim 1/p$ for $p \to \infty$. See ref.~\cite{QCDT0} for further discussion on the vector kernels.

Finally, fig.~\ref{Abb: Besetzungszahl} shows the occupation number density of quark states \cite{QCDT0}\footnote{Note that there is no summation over spin and color indices on the l.h.s.}
\beq
\frac{\langle {a^{s, m}}^{\dagger}(\vp) a^{s, m}(\vp) \rangle}{(2 \pi)^3 \delta^3(0)} = P(p) S^2(p) \, . \label{Gl: Besetzungszahl}
\eeq
On a linear scale, the results for $g = 2.1$ and $g = 0$ are almost indistinguishable. Note that for the chosen ansatz for the vacuum wave functional the densities of occupied quark and anti-quark states agree.  

\section{Conclusions} \label{Abschn: Zusammenfassung}

In this paper, we have carried out a variational calculation within the Hamilton approach to QCD \cite{QCDT0}. The vacuum wave functional used includes the quark-gluon coupling with two different Dirac structures. The vacuum energy is calculated up to including two-loop order. In the resulting gap equation the linear UV divergences induced by these two Dirac structures cancel. When, in addition, the Coulomb potential with its correct UV form is included, also the logarithmic UV divergences cancel. The resulting finite variational equations were solved numerically. When the Coulomb string tension is put to zero, chiral symmetry turns out to be not spontaneously broken. Assuming a Coulomb string tension of $\sigma_{\mathrm{C}} = 2.5 \sigma$ with $\sigma$ being the Wilsonian string tension the phenomenological value of the quark condensate $\langle \bar{\psi} \psi \rangle \simeq (- 235 \, \mathrm{MeV})^3$ was reproduced for a value of $g \simeq 2.1$ of the quark-gluon coupling constant.

The variational solution of QCD obtained in the present paper will serve as input in a forthcoming investigation of the chiral and deconfinement phase transitions. 

\section*{Acknowledgments}

The authors thank M.~Quandt and H.~Vogt for helpful discussions. This work was supported by Deutsche Forschungsgemeinschaft (DFG) under Contract No.~DFG-Re856/10-1.

\bibliography{QCDT0}

\end{document}